
%
%
%
%
%
%
%
\def\standardrisposta{s }\def\reducedrisposta{r }
\def\mplarisposta{mpla }
\def\doublerisposta{d }\def\cartarisposta{e }\def\amsrisposta{y }
\newcount\ingrandimento \newcount\sinnota \newcount\dimnota
\newcount\unoduecol \newdimen\collhsize \newdimen\tothsize
\newdimen\fullhsize \newcount\controllorisposta \sinnota=1
\newskip\infralinea  \global\controllorisposta=0
\message{ ********    Welcome to PANDA macros (Plain TeX, AP, 1991)}
\message{ ******** }
\message{       You'll have to answer a few questions in lowercase.}
\message{>  Do you want it in double-page (d), reduced (r)}
\message{or standard format (s) ? }\read-1 to\risposta
\message{>  Do you want it in USA A4 (u) or EUROPEAN A4 (e)}
\message{paper size ? }\read-1 to\srisposta
\message{>  Do you have AMSFonts 2.0 (math) fonts (y/n) ? }
\read-1 to\arisposta
%
%
%
%
%
\ifx\risposta\standardrisposta \ingrandimento=1200
\message{>> This will come out UNREDUCED << }
\dimnota=2 \unoduecol=1 \global\controllorisposta=1 \fi
\ifx\risposta\reducedrisposta \ingrandimento=1095 \dimnota=1
\unoduecol=1  \global\controllorisposta=1
\message{>> This will come out REDUCED << } \fi
\ifx\risposta\doublerisposta \ingrandimento=1000 \dimnota=2
\unoduecol=2  \global\controllorisposta=1 
\message{>> You must print this in LANDSCAPE orientation << } \fi
\ifx\risposta\mplarisposta \ingrandimento=1000 \dimnota=1
\message{>> Mod. Phys. Lett. A format << }
\unoduecol=1 \global\controllorisposta=1 \fi
\ifnum\controllorisposta=0  \ingrandimento=1200
\message{>>> ERROR IN INPUT, I ASSUME STANDARD UNREDUCED FORMAT <<< }
\dimnota=2 \unoduecol=1 \fi
\magnification=\ingrandimento
%
%
%
%
\newdimen\eucolumnsize \newdimen\eudoublehsize \newdimen\eudoublevsize
\newdimen\uscolumnsize \newdimen\usdoublehsize \newdimen\usdoublevsize
\newdimen\eusinglehsize \newdimen\eusinglevsize \newdimen\ussinglehsize
\newskip\standardbaselineskip \newdimen\ussinglevsize
\newskip\reducedbaselineskip \newskip\doublebaselineskip
\eucolumnsize=12.0truecm    
\eudoublehsize=25.5truecm   
\eudoublevsize=6.5truein    
\uscolumnsize=4.4truein     
\usdoublehsize=9.4truein    
\usdoublevsize=6.8truein    
\eusinglehsize=6.5truein    
\eusinglevsize=24truecm     
\ussinglehsize=6.5truein    
\ussinglevsize=8.9truein    
\standardbaselineskip=16pt  
\reducedbaselineskip=14pt   
\doublebaselineskip=12pt    
%
%
\def\Portoffset{}
\def\Landoffset{}
\ifx\risposta\mplarisposta \def\Portoffset{\hoffset=1.8truecm} \fi
%
%
\def\Landspec{}
\tolerance=10000
\parskip 0pt plus 2pt  \leftskip=0pt \rightskip=0pt
%
%
%
\ifx\risposta\standardrisposta \infralinea=\standardbaselineskip \fi
\ifx\risposta\reducedrisposta  \infralinea=\reducedbaselineskip \fi
\ifx\risposta\doublerisposta   \infralinea=\doublebaselineskip \fi
\ifx\risposta\mplarisposta     \infralinea=13pt \fi
\ifnum\controllorisposta=0    \infralinea=\standardbaselineskip \fi
\ifx\risposta\doublerisposta   \Landoffset \else \Portoffset \fi
\ifx\risposta\doublerisposta \ifx\srisposta\cartarisposta
\tothsize=\eudoublehsize \collhsize=\eucolumnsize
\vsize=\eudoublevsize  \else  \tothsize=\usdoublehsize
\collhsize=\uscolumnsize \vsize=\usdoublevsize \fi \else
\ifx\srisposta\cartarisposta \tothsize=\eusinglehsize
\vsize=\eusinglevsize \else  \tothsize=\ussinglehsize
\vsize=\ussinglevsize \fi \collhsize=4.4truein \fi
\ifx\risposta\mplarisposta \tothsize=5.0truein
\vsize=7.8truein \collhsize=4.4truein \fi
%
%
%
%
\newcount\contaeuler \newcount\contacyrill \newcount\contaams
\font\ninerm=cmr9  \font\eightrm=cmr8  \font\sixrm=cmr6
\font\ninei=cmmi9  \font\eighti=cmmi8  \font\sixi=cmmi6
\font\ninesy=cmsy9  \font\eightsy=cmsy8  \font\sixsy=cmsy6
\font\ninebf=cmbx9  \font\eightbf=cmbx8  \font\sixbf=cmbx6
\font\ninett=cmtt9  \font\eighttt=cmtt8  \font\nineit=cmti9
\font\eightit=cmti8 \font\ninesl=cmsl9  \font\eightsl=cmsl8
\skewchar\ninei='177 \skewchar\eighti='177 \skewchar\sixi='177
\skewchar\ninesy='60 \skewchar\eightsy='60 \skewchar\sixsy='60
\hyphenchar\ninett=-1 \hyphenchar\eighttt=-1 \hyphenchar\tentt=-1
%
\font\tencmmib=cmmib10  \newfam\cmmibfam  \skewchar\tencmmib='177
\font\tencmbsy=cmbsy10  \newfam\cmbsyfam  \skewchar\tencmbsy='60
\def\scaps{\cmcsc}                 
\font\tencmcsc=cmcsc10  \newfam\cmcscfam
\ifnum\ingrandimento=1095

\font\capsone=cmcsc10 at 10.95pt 

\else

\font\capsone=cmcsc10 at 12pt 
\fi

\def\ttaarr{\bf}		
\def\ppaarr{\sl}		

%
%
%
\newfam\eufmfam \newfam\msamfam \newfam\msbmfam \newfam\eufbfam
\def\Loadeulerfonts{\global\contaeuler=1 \ifx\arisposta\amsrisposta
\font\teneufm=eufm10              
\font\eighteufm=eufm8 \font\nineeufm=eufm9 \font\sixeufm=eufm6
\font\seveneufm=eufm7  \font\fiveeufm=eufm5
\font\teneufb=eufb10              
\font\eighteufb=eufb8 \font\nineeufb=eufb9 \font\sixeufb=eufb6
\font\seveneufb=eufb7  \font\fiveeufb=eufb5
\font\teneurm=eurm10              
\font\eighteurm=eurm8 \font\nineeurm=eurm9
\font\teneurb=eurb10              
\font\eighteurb=eurb8 \font\nineeurb=eurb9
\font\teneusm=eusm10              
\font\eighteusm=eusm8 \font\nineeusm=eusm9
\font\teneusb=eusb10              
\font\eighteusb=eusb8 \font\nineeusb=eusb9
\else \def\eufm{\tt} \def\eufb{\tt} \def\eurm{\tt} \def\eurb{\tt}
\def\eusm{\tt} \def\eusb{\tt}    \fi}

\def\loadamsmath{\global\contaams=1 \ifx\arisposta\amsrisposta
\font\tenmsam=msam10 \font\ninemsam=msam9 \font\eightmsam=msam8
\font\sevenmsam=msam7 \font\sixmsam=msam6 \font\fivemsam=msam5
\font\tenmsbm=msbm10 \font\ninemsbm=msbm9 \font\eightmsbm=msbm8
\font\sevenmsbm=msbm7 \font\sixmsbm=msbm6 \font\fivemsbm=msbm5
\else \def\msbm{\bf} \fi \def\Bbb{\msbm} \def\symbl{\msam} \tenpoint}
\def\loadcyrill{\global\contacyrill=1 \ifx\arisposta\amsrisposta
\font\tenwncyr=wncyr10 \font\ninewncyr=wncyr9 \font\eightwncyr=wncyr8
\font\tenwncyb=wncyr10 \font\ninewncyb=wncyr9 \font\eightwncyb=wncyr8
\font\tenwncyi=wncyr10 \font\ninewncyi=wncyr9 \font\eightwncyi=wncyr8
\else \def\cyrill{\sl} \def\cyrilb{\sl} \def\cyrili{\sl} \fi\tenpoint}
\ifx\arisposta\amsrisposta
\font\sevenex=cmex7               
\font\eightex=cmex8  \font\nineex=cmex9
\font\ninecmmib=cmmib9   \font\eightcmmib=cmmib8
\font\sevencmmib=cmmib7 \font\sixcmmib=cmmib6
\font\fivecmmib=cmmib5   \skewchar\ninecmmib='177
\skewchar\eightcmmib='177  \skewchar\sevencmmib='177
\skewchar\sixcmmib='177   \skewchar\fivecmmib='177
\font\ninecmbsy=cmbsy9    \font\eightcmbsy=cmbsy8
\font\sevencmbsy=cmbsy7  \font\sixcmbsy=cmbsy6
\font\fivecmbsy=cmbsy5   \skewchar\ninecmbsy='60
\skewchar\eightcmbsy='60  \skewchar\sevencmbsy='60
\skewchar\sixcmbsy='60    \skewchar\fivecmbsy='60
\font\ninecmcsc=cmcsc9    \font\eightcmcsc=cmcsc8     \else
\def\cmmib{\fam\cmmibfam\tencmmib}\textfont\cmmibfam=\tencmmib
\scriptfont\cmmibfam=\tencmmib \scriptscriptfont\cmmibfam=\tencmmib
\def\cmbsy{\fam\cmbsyfam\tencmbsy} \textfont\cmbsyfam=\tencmbsy
\scriptfont\cmbsyfam=\tencmbsy \scriptscriptfont\cmbsyfam=\tencmbsy
\scriptfont\cmcscfam=\tencmcsc \scriptscriptfont\cmcscfam=\tencmcsc
\def\cmcsc{\fam\cmcscfam\tencmcsc} \textfont\cmcscfam=\tencmcsc \fi
\catcode`@=11
\newskip\ttglue
\gdef\tenpoint{\def\rm{\fam0\tenrm}
  \textfont0=\tenrm \scriptfont0=\sevenrm \scriptscriptfont0=\fiverm
  \textfont1=\teni \scriptfont1=\seveni \scriptscriptfont1=\fivei
  \textfont2=\tensy \scriptfont2=\sevensy \scriptscriptfont2=\fivesy
  \textfont3=\tenex \scriptfont3=\tenex \scriptscriptfont3=\tenex
  \def\mcal{\fam2 \tensy}  \def\mmit{\fam1 \teni}
  \textfont\itfam=\tenit \def\it{\fam\itfam\tenit}
  \textfont\slfam=\tensl \def\sl{\fam\slfam\tensl}
  \textfont\ttfam=\tentt \scriptfont\ttfam=\eighttt
  \scriptscriptfont\ttfam=\eighttt  \def\tt{\fam\ttfam\tentt}
  \textfont\bffam=\tenbf \scriptfont\bffam=\sevenbf
  \scriptscriptfont\bffam=\fivebf \def\bf{\fam\bffam\tenbf}
     \ifx\arisposta\amsrisposta    \ifnum\contaeuler=1
  \textfont\eufmfam=\teneufm \scriptfont\eufmfam=\seveneufm
  \scriptscriptfont\eufmfam=\fiveeufm \def\eufm{\fam\eufmfam\teneufm}
  \textfont\eufbfam=\teneufb \scriptfont\eufbfam=\seveneufb
  \scriptscriptfont\eufbfam=\fiveeufb \def\eufb{\fam\eufbfam\teneufb}
  \def\eurm{\teneurm} \def\eurb{\teneurb} \def\eusm{\teneusm}
  \def\eusb{\teneusb}    \fi    \ifnum\contaams=1
  \textfont\msamfam=\tenmsam \scriptfont\msamfam=\sevenmsam
  \scriptscriptfont\msamfam=\fivemsam \def\msam{\fam\msamfam\tenmsam}
  \textfont\msbmfam=\tenmsbm \scriptfont\msbmfam=\sevenmsbm
  \scriptscriptfont\msbmfam=\fivemsbm \def\msbm{\fam\msbmfam\tenmsbm}
     \fi      \ifnum\contacyrill=1     \def\cyrill{\tenwncyr}
  \def\cyrilb{\tenwncyb}  \def\cyrili{\tenwncyi}         \fi
  \textfont3=\tenex \scriptfont3=\sevenex \scriptscriptfont3=\sevenex
  \def\cmmib{\fam\cmmibfam\tencmmib} \scriptfont\cmmibfam=\sevencmmib
  \textfont\cmmibfam=\tencmmib  \scriptscriptfont\cmmibfam=\fivecmmib
  \def\cmbsy{\fam\cmbsyfam\tencmbsy} \scriptfont\cmbsyfam=\sevencmbsy
  \textfont\cmbsyfam=\tencmbsy  \scriptscriptfont\cmbsyfam=\fivecmbsy
  \def\cmcsc{\fam\cmcscfam\tencmcsc} \scriptfont\cmcscfam=\eightcmcsc
  \textfont\cmcscfam=\tencmcsc \scriptscriptfont\cmcscfam=\eightcmcsc
     \fi            \tt \ttglue=.5em plus.25em minus.15em
  \normalbaselineskip=12pt
  \setbox\strutbox=\hbox{\vrule height8.5pt depth3.5pt width0pt}
  \let\sc=\eightrm \let\big=\tenbig   \normalbaselines
  \baselineskip=\infralinea  \rm}
\gdef\ninepoint{\def\rm{\fam0\ninerm}
  \textfont0=\ninerm \scriptfont0=\sixrm \scriptscriptfont0=\fiverm
  \textfont1=\ninei \scriptfont1=\sixi \scriptscriptfont1=\fivei
  \textfont2=\ninesy \scriptfont2=\sixsy \scriptscriptfont2=\fivesy
  \textfont3=\tenex \scriptfont3=\tenex \scriptscriptfont3=\tenex
  \def\mcal{\fam2 \ninesy}  \def\mmit{\fam1 \ninei}
  \textfont\itfam=\nineit \def\it{\fam\itfam\nineit}
  \textfont\slfam=\ninesl \def\sl{\fam\slfam\ninesl}
  \textfont\ttfam=\ninett \scriptfont\ttfam=\eighttt
  \scriptscriptfont\ttfam=\eighttt \def\tt{\fam\ttfam\ninett}
  \textfont\bffam=\ninebf \scriptfont\bffam=\sixbf
  \scriptscriptfont\bffam=\fivebf \def\bf{\fam\bffam\ninebf}
     \ifx\arisposta\amsrisposta  \ifnum\contaeuler=1
  \textfont\eufmfam=\nineeufm \scriptfont\eufmfam=\sixeufm
  \scriptscriptfont\eufmfam=\fiveeufm \def\eufm{\fam\eufmfam\nineeufm}
  \textfont\eufbfam=\nineeufb \scriptfont\eufbfam=\sixeufb
  \scriptscriptfont\eufbfam=\fiveeufb \def\eufb{\fam\eufbfam\nineeufb}
  \def\eurm{\nineeurm} \def\eurb{\nineeurb} \def\eusm{\nineeusm}
  \def\eusb{\nineeusb}     \fi   \ifnum\contaams=1
  \textfont\msamfam=\ninemsam \scriptfont\msamfam=\sixmsam
  \scriptscriptfont\msamfam=\fivemsam \def\msam{\fam\msamfam\ninemsam}
  \textfont\msbmfam=\ninemsbm \scriptfont\msbmfam=\sixmsbm
  \scriptscriptfont\msbmfam=\fivemsbm \def\msbm{\fam\msbmfam\ninemsbm}
     \fi       \ifnum\contacyrill=1     \def\cyrill{\ninewncyr}
  \def\cyrilb{\ninewncyb}  \def\cyrili{\ninewncyi}         \fi
  \textfont3=\nineex \scriptfont3=\sevenex \scriptscriptfont3=\sevenex
  \def\cmmib{\fam\cmmibfam\ninecmmib}  \textfont\cmmibfam=\ninecmmib
  \scriptfont\cmmibfam=\sixcmmib \scriptscriptfont\cmmibfam=\fivecmmib
  \def\cmbsy{\fam\cmbsyfam\ninecmbsy}  \textfont\cmbsyfam=\ninecmbsy
  \scriptfont\cmbsyfam=\sixcmbsy \scriptscriptfont\cmbsyfam=\fivecmbsy
  \def\cmcsc{\fam\cmcscfam\ninecmcsc} \scriptfont\cmcscfam=\eightcmcsc
  \textfont\cmcscfam=\ninecmcsc \scriptscriptfont\cmcscfam=\eightcmcsc
     \fi            \tt \ttglue=.5em plus.25em minus.15em
  \normalbaselineskip=11pt
  \setbox\strutbox=\hbox{\vrule height8pt depth3pt width0pt}
  \let\sc=\sevenrm \let\big=\ninebig \normalbaselines\rm}
\gdef\eightpoint{\def\rm{\fam0\eightrm}
  \textfont0=\eightrm \scriptfont0=\sixrm \scriptscriptfont0=\fiverm
  \textfont1=\eighti \scriptfont1=\sixi \scriptscriptfont1=\fivei
  \textfont2=\eightsy \scriptfont2=\sixsy \scriptscriptfont2=\fivesy
  \textfont3=\tenex \scriptfont3=\tenex \scriptscriptfont3=\tenex
  \def\mcal{\fam2 \eightsy}  \def\mmit{\fam1 \eighti}
  \textfont\itfam=\eightit \def\it{\fam\itfam\eightit}
  \textfont\slfam=\eightsl \def\sl{\fam\slfam\eightsl}
  \textfont\ttfam=\eighttt \scriptfont\ttfam=\eighttt
  \scriptscriptfont\ttfam=\eighttt \def\tt{\fam\ttfam\eighttt}
  \textfont\bffam=\eightbf \scriptfont\bffam=\sixbf
  \scriptscriptfont\bffam=\fivebf \def\bf{\fam\bffam\eightbf}
     \ifx\arisposta\amsrisposta   \ifnum\contaeuler=1
  \textfont\eufmfam=\eighteufm \scriptfont\eufmfam=\sixeufm
  \scriptscriptfont\eufmfam=\fiveeufm \def\eufm{\fam\eufmfam\eighteufm}
  \textfont\eufbfam=\eighteufb \scriptfont\eufbfam=\sixeufb
  \scriptscriptfont\eufbfam=\fiveeufb \def\eufb{\fam\eufbfam\eighteufb}
  \def\eurm{\eighteurm} \def\eurb{\eighteurb} \def\eusm{\eighteusm}
  \def\eusb{\eighteusb}       \fi    \ifnum\contaams=1
  \textfont\msamfam=\eightmsam \scriptfont\msamfam=\sixmsam
  \scriptscriptfont\msamfam=\fivemsam \def\msam{\fam\msamfam\eightmsam}
  \textfont\msbmfam=\eightmsbm \scriptfont\msbmfam=\sixmsbm
  \scriptscriptfont\msbmfam=\fivemsbm \def\msbm{\fam\msbmfam\eightmsbm}
     \fi       \ifnum\contacyrill=1     \def\cyrill{\eightwncyr}
  \def\cyrilb{\eightwncyb}  \def\cyrili{\eightwncyi}         \fi
  \textfont3=\eightex \scriptfont3=\sevenex \scriptscriptfont3=\sevenex
  \def\cmmib{\fam\cmmibfam\eightcmmib}  \textfont\cmmibfam=\eightcmmib
  \scriptfont\cmmibfam=\sixcmmib \scriptscriptfont\cmmibfam=\fivecmmib
  \def\cmbsy{\fam\cmbsyfam\eightcmbsy}  \textfont\cmbsyfam=\eightcmbsy
  \scriptfont\cmbsyfam=\sixcmbsy \scriptscriptfont\cmbsyfam=\fivecmbsy
  \def\cmcsc{\fam\cmcscfam\eightcmcsc} \scriptfont\cmcscfam=\eightcmcsc
  \textfont\cmcscfam=\eightcmcsc \scriptscriptfont\cmcscfam=\eightcmcsc
     \fi             \tt \ttglue=.5em plus.25em minus.15em
  \normalbaselineskip=9pt
  \setbox\strutbox=\hbox{\vrule height7pt depth2pt width0pt}
  \let\sc=\sixrm \let\big=\eightbig \normalbaselines\rm }
\gdef\tenbig#1{{\hbox{$\left#1\vbox to8.5pt{}\right.\n@space$}}}
\gdef\ninebig#1{{\hbox{$\textfont0=\tenrm\textfont2=\tensy
   \left#1\vbox to7.25pt{}\right.\n@space$}}}
\gdef\eightbig#1{{\hbox{$\textfont0=\ninerm\textfont2=\ninesy
   \left#1\vbox to6.5pt{}\right.\n@space$}}}
\def\alternativefont#1#2{\ifx\arisposta\amsrisposta \relax \else
\xdef#1{#2} \fi}
\global\contaeuler=0 \global\contacyrill=0 \global\contaams=0
%
%
%
%
\newbox\fotlinebb \newbox\hedlinebb \newbox\leftcolumn
\gdef\makeheadline{\vbox to 0pt{\vskip-22.5pt
     \fullline{\vbox to8.5pt{}\the\headline}\vss}\nointerlineskip}
\gdef\makehedlinebb{\vbox to 0pt{\vskip-22.5pt
     \fullline{\vbox to8.5pt{}\copy\hedlinebb\hfil
     \line{\hfill\the\headline\hfill}}\vss} \nointerlineskip}
\gdef\makefootline{\baselineskip=24pt \fullline{\the\footline}}
\gdef\makefotlinebb{\baselineskip=24pt
    \fullline{\copy\fotlinebb\hfil\line{\hfill\the\footline\hfill}}}
\gdef\doubleformat{\shipout\vbox{\Landspec\makehedlinebb
     \fullline{\box\leftcolumn\hfil\columnbox}\makefotlinebb}
     \advancepageno}
\gdef\columnbox{\leftline{\pagebody}}
\gdef\line#1{\hbox to\hsize{\hskip\leftskip#1\hskip\rightskip}}
\gdef\fullline#1{\hbox to\fullhsize{\hskip\leftskip{#1}%
\hskip\rightskip}}
\gdef\footnote#1{\let\@sf=\empty
         \ifhmode\edef\#sf{\spacefactor=\the\spacefactor}\/\fi
         #1\@sf\vfootnote{#1}}
\gdef\vfootnote#1{\insert\footins\bgroup
         \ifnum\dimnota=1  \eightpoint\fi
         \ifnum\dimnota=2  \ninepoint\fi
         \ifnum\dimnota=0  \tenpoint\fi
         \interlinepenalty=\interfootnotelinepenalty
         \splittopskip=\ht\strutbox
         \splitmaxdepth=\dp\strutbox \floatingpenalty=20000
         \leftskip=\oldssposta \rightskip=\olddsposta
         \spaceskip=0pt \xspaceskip=0pt
         \ifnum\sinnota=0   \textindent{#1}\fi
         \ifnum\sinnota=1   \item{#1}\fi
         \footstrut\futurelet\next\fo@t}
\gdef\fo@t{\ifcat\bgroup\noexpand\next \let\next\f@@t
             \else\let\next\f@t\fi \next}
\gdef\f@@t{\bgroup\aftergroup\@foot\let\next}
\gdef\f@t#1{#1\@foot} \gdef\@foot{\strut\egroup}
\gdef\footstrut{\vbox to\splittopskip{}}
\skip\footins=\bigskipamount
\count\footins=1000  \dimen\footins=8in
\catcode`@=12
\tenpoint
\ifnum\unoduecol=1 \hsize=\tothsize   \fullhsize=\tothsize \fi
\ifnum\unoduecol=2 \hsize=\collhsize  \fullhsize=\tothsize \fi
\global\let\lrcol=L
\ifnum\unoduecol=1 \output{\plainoutput{\ifnum\tipbnota=2
\clearnmbnota\fi}} \fi
\ifnum\unoduecol=2 \output{\if L\lrcol
     \global\setbox\leftcolumn=\columnbox
     \global\setbox\fotlinebb=\line{\hfill\the\footline\hfill}
     \global\setbox\hedlinebb=\line{\hfill\the\headline\hfill}
     \advancepageno  \global\let\lrcol=R
     \else  \doubleformat \global\let\lrcol=L \fi
     \ifnum\outputpenalty>-20000 \else\dosupereject\fi
     \ifnum\tipbnota=2\clearnmbnota\fi }\fi
\def\ifdoublepage{\ifnum\unoduecol=2 }
\gdef\yespagenumbers{\footline={\hss\tenrm\folio\hss}}
\gdef\ciao{\par\vfill\supereject \ifnum\unoduecol=2
     \if R\lrcol  \headline={}\nopagenumbers\null\vfill\eject
     \fi\fi \end}

\newskip\olddsposta \newskip\oldssposta
\global\oldssposta=\leftskip \global\olddsposta=\rightskip

\def\filldots{\leaders\hbox to 1em{\hss.\hss}\hfill}
\def\inquadrb#1 {\vbox {\hrule  \hbox{\vrule \vbox {\vskip .2cm
    \hbox {\ #1\ } \vskip .2cm } \vrule  }  \hrule} }
 \def\newline{\hfil\break}
\def\jump{\vskip\baselineskip} \newskip\iinnffrr
\def\sjump{\iinnffrr=\baselineskip
          \divide\iinnffrr by 2 \vskip\iinnffrr}
\def\bjump{\vskip\baselineskip \vskip\baselineskip}
\newcount\nmbnota  \def\clearnmbnota{\global\nmbnota=0}
\newcount\tipbnota \def\letterfootnote{\global\tipbnota=1}

\def\note#1{\global\advance\nmbnota by 1 \ifnum\tipbnota=1
    \footnote{$^{\rm\nttlett}$}{#1} \else {\ifnum\tipbnota=2
    \footnote{$^{\nttsymb}$}{#1}
    \else\footnote{$^{\the\nmbnota}$}{#1}\fi}\fi}
\def\nttlett{\ifcase\nmbnota \or a\or b\or c\or d\or e\or f\or
g\or h\or i\or j\or k\or l\or m\or n\or o\or p\or q\or r\or
s\or t\or u\or v\or w\or y\or x\or z\fi}
\def\nttsymb{\ifcase\nmbnota \or\dag\or\sharp\or\ddag\or\star\or
\natural\or\flat\or\clubsuit\or\diamondsuit\or\heartsuit
\or\spadesuit\fi}   \clearnmbnota
\def\numberfootnote{\global\tipbnota=0} \numberfootnote
\def\setnote#1{\expandafter\xdef\csname#1\endcsname{
\ifnum\tipbnota=1 {\rm\nttlett} \else {\ifnum\tipbnota=2
{\nttsymb} \else \the\nmbnota\fi}\fi} }
\newcount\nbmfig  \def\clearnbmfig{\global\nbmfig=0}
\gdef\figure{\global\advance\nbmfig by 1
      {\rm fig. \the\nbmfig}}   \clearnbmfig
\def\setfig#1{\expandafter\xdef\csname#1\endcsname{fig. \the\nbmfig}}

\newcount\frmcount \def\clearfrmcount{\global\frmcount=0}
\def\numero{\global\advance\frmcount by 1   \ifnum\indappcount=0
  {\ifnum\cpcount <1 {\hbox{\rm (\the\frmcount )}}  \else
  {\hbox{\rm (\the\cpcount .\the\frmcount )}} \fi}  \else
  {\hbox{\rm (\applett .\the\frmcount )}} \fi}
\def\nameformula#1{\global\advance\frmcount by 1%
\ifnum\draftnum=0  {\ifnum\indappcount=0%
{\ifnum\cpcount<1\xdef\spzzttrra{(\the\frmcount )}%
\else\xdef\spzzttrra{(\the\cpcount .\the\frmcount )}\fi}%
\else\xdef\spzzttrra{(\applett .\the\frmcount )}\fi}%
\else\xdef\spzzttrra{(#1)}\fi%
\expandafter\xdef\csname#1\endcsname{\spzzttrra}
\eqno \ifnum\draftnum=0 {\ifnum\indappcount=0
  {\ifnum\cpcount <1 {\hbox{\rm (\the\frmcount )}}  \else
  {\hbox{\rm (\the\cpcount .\the\frmcount )}}\fi}   \else
  {\hbox{\rm (\applett .\the\frmcount )}} \fi} \else (#1) \fi $$}
\def\nfr{\nameformula}    
\def\nameali#1{\global\advance\frmcount by 1%
\ifnum\draftnum=0  {\ifnum\indappcount=0%
{\ifnum\cpcount<1\xdef\spzzttrra{(\the\frmcount )}%
\else\xdef\spzzttrra{(\the\cpcount .\the\frmcount )}\fi}%
\else\xdef\spzzttrra{(\applett .\the\frmcount )}\fi}%
\else\xdef\spzzttrra{(#1)}\fi%
\expandafter\xdef\csname#1\endcsname{\spzzttrra}
  \ifnum\draftnum=0  {\ifnum\indappcount=0
  {\ifnum\cpcount <1 {\hbox{\rm (\the\frmcount )}}  \else
  {\hbox{\rm (\the\cpcount .\the\frmcount )}}\fi}   \else
  {\hbox{\rm (\applett .\the\frmcount )}} \fi} \else (#1) \fi}
\clearfrmcount
\newcount\cpcount \def\clearcpcount{\global\cpcount=0}
\newcount\subcpcount \def\clearsubcpcount{\global\subcpcount=0}
\newcount\appcount \def\clearappcount{\global\appcount=0}
\newcount\indappcount \def\clearindappcount{\indappcount=0}
\newcount\sottoparcount 

\def\applett{\ifcase\appcount  \or {A}\or {B}\or {C}\or
{D}\or {E}\or {F}\or {G}\or {H}\or {I}\or {J}\or {K}\or {L}\or
{M}\or {N}\or {O}\or {P}\or {Q}\or {R}\or {S}\or {T}\or {U}\or
{V}\or {W}\or {X}\or {Y}\or {Z}\fi
             \ifnum\appcount<0
    \message{>>  ERROR: counter \appcount out of range <<}\fi
             \ifnum\appcount>26
   \message{>>  ERROR: counter \appcount out of range <<}\fi}
\clearappcount  \clearindappcount
\newcount\connttrre  \def\clearconnttrre{\global\connttrre=0}
\newcount\countref  \def\clearcountref{\global\countref=0}
\clearcountref
\def\chapter#1{\global\advance\cpcount by 1 \clearfrmcount
                 \goodbreak\null\vbox{\jump\nobreak
                 \clearsubcpcount\clearindappcount
                 \itemitem{\ttaarr\the\cpcount .\qquad}{\ttaarr #1}
                 \par\nobreak\jump\sjump}\nobreak}
\def\section#1{\global\advance\subcpcount by 1 \goodbreak\null
               \vbox{\sjump\nobreak\ifnum\indappcount=0
                 {\ifnum\cpcount=0 {\itemitem{\ppaarr
               .\the\subcpcount\quad\enskip\ }{\ppaarr #1}\par} \else
                 {\itemitem{\ppaarr\the\cpcount .\the\subcpcount\quad
                  \enskip\ }{\ppaarr #1} \par}  \fi}
                \else{\itemitem{\ppaarr\applett .\the\subcpcount\quad
                 \enskip\ }{\ppaarr #1}\par}\fi\nobreak\jump}\nobreak}
\clearsubcpcount
\def\appendix#1{\global\advance\appcount by 1 \clearfrmcount
                  \goodbreak\null\vbox{\jump\nobreak
                  \global\advance\indappcount by 1 \clearsubcpcount
                  \itemitem{\ttaarr App.\applett\ }{\ttaarr #1}
                  \nobreak\jump\sjump}\nobreak}
\clearappcount \clearindappcount

\clearcpcount\clearcountref

\def\setchap#1{\ifnum\indappcount=0{\ifnum\subcpcount=0%
\xdef\spzzttrra{\the\cpcount}%
\else\xdef\spzzttrra{\the\cpcount .\the\subcpcount}\fi}
\else{\ifnum\subcpcount=0 \xdef\spzzttrra{\applett}%
\else\xdef\spzzttrra{\applett .\the\subcpcount}\fi}\fi
\expandafter\xdef\csname#1\endcsname{\spzzttrra}}
\newcount\draftnum \newcount\ppora   \newcount\ppminuti
\global\ppora=\time   \global\ppminuti=\time
\global\divide\ppora by 60  \draftnum=\ppora
\multiply\draftnum by 60    \global\advance\ppminuti by -\draftnum
\global\draftnum=0
\def\droggi{\number\day /\number\month /\number\year\ \the\ppora
:\the\ppminuti}
 \global\draftnum=0
\def\draftcomment#1{\ifnum\draftnum=0 \relax \else
{\ {\bf ***}\ #1\ {\bf ***}\ }\fi} 
%
%
\catcode`@=11
\gdef\Ref#1{\expandafter\ifx\csname @rrxx@#1\endcsname\relax%
{\global\advance\countref by 1%
\ifnum\countref>200%
\message{>>> ERROR: maximum number of references exceeded <<<}%
\expandafter\xdef\csname @rrxx@#1\endcsname{0}\else%
\expandafter\xdef\csname @rrxx@#1\endcsname{\the\countref}\fi}\fi%
\ifnum\draftnum=0 \csname @rrxx@#1\endcsname \else#1\fi}
\gdef\beginref{\ifnum\draftnum=0  \gdef\Rref{\fairef}
\gdef\endref{\scriviref} \else\relax\fi
\ifx\risposta\mplarisposta \ninepoint \fi
\parskip 2pt plus 2pt \baselineskip=12pt}
\def\Reflab#1{[#1]} \gdef\Rref#1#2{\item{\Reflab{#1}}{#2}}
\gdef\endref{\relax}  \newcount\conttemp
\gdef\fairef#1#2{\expandafter\ifx\csname @rrxx@#1\endcsname\relax
{\global\conttemp=0
\message{>>> ERROR: reference [#1] not defined <<<} } \else
{\global\conttemp=\csname @rrxx@#1\endcsname } \fi
\global\advance\conttemp by 50
\global\setbox\conttemp=\hbox{#2} }
\gdef\scriviref{\clearconnttrre\conttemp=50
\loop\ifnum\connttrre<\countref \advance\conttemp by 1
\advance\connttrre by 1
\item{\Reflab{\the\connttrre}}{\unhcopy\conttemp} \repeat}
\clearcountref \clearconnttrre
\catcode`@=12
\ifx\risposta\mplarisposta \def\Reflab#1{#1.} \letterfootnote \fi

\def\slashchar#1{\setbox0=\hbox{$#1$} \dimen0=\wd0
     \setbox1=\hbox{/} \dimen1=\wd1 \ifdim\dimen0>\dimen1
      \rlap{\hbox to \dimen0{\hfil/\hfil}} #1 \else
      \rlap{\hbox to \dimen1{\hfil$#1$\hfil}} / \fi}
\ifx\oldchi\undefined \let\oldchi=\chi
  \def\cchi{{\raise 1pt\hbox{$\oldchi$}}} \let\chi=\cchi \fi
  \def\grad{\nabla}

\def\frac#1#2{{\textstyle{#1 \over #2}}}

\def\half{\ifinner {\scriptstyle {1 \over 2}}\else {1 \over 2} \fi}

\def\simge{\rlap{\raise 2pt \hbox{$>$}}{\lower 2pt \hbox{$\sim$}}}
\def\simle{\rlap{\raise 2pt \hbox{$<$}}{\lower 2pt \hbox{$\sim$}}}

\def\vbig#1#2{{\vbigd@men=#2\divide\vbigd@men by 2%
\hbox{$\left#1\vbox to \vbigd@men{}\right.\n@space$}}}

\null
%
%
%
%

\nopagenumbers{\baselineskip=12pt
\line{\hfill CBPF-NF-057194/CLAF.}
\ifdoublepage \bjump\bjump\bjump\bjump\else\vfill\fi
\centerline{\capsone The BV Quantisation of Superparticles}
\sjump
\centerline{\capsone Type I and II.}
\bjump
\centerline{\scaps {Jos\'e-Luis V\'azquez-Bello}}
\sjump
\centerline{\sl CBPF-CNPq/CLAF Centro Brasileiro de Pesquisas Fisicas}
\centerline{\sl Rua Dr. Xavier Sigaud 150, CEP. 22290}
\centerline{\sl  Rio de  Janeiro - RJ, BRASIL.}
\sjump
\centerline{ {\sl e-mail:} bello@cbpfsu7.cat.cbpf.br}
\vfill
\ifnum\unoduecol=2 \eject\null\vfill\fi

\centerline{\capsone abstract }
\sjump
\noindent

{ This letter discusses the BRST cohomology of superparticles type I and II.
It was used an extended super-space to construct $S0(9,1)$ superparticle
actions that lead to super-wave functions whose spinor components
satisfy $S0(9,1)$
covariant constraints. Their BRST charges were found by using BV methods,
since the models present a large number of symmetries and only close on-shell.
It is shown that the zero ghost-number cohomology class of both models
reproduce the same spectrum as that of N=1 ten dimensional super-Yang-Mills
theory. }

\sjump
\ifnum\unoduecol=2 \vfill\fi
\eject

\yespagenumbers\pageno=1
%

\def\psh {\rlap{/}{p}}

\def\Lamdsh {\Lambda\llap{/}}

\def\Sigmash {\Sigma\llap{/}}

\def\xmu { x^\mu}
\def\pmu { p_\mu}
\def\half {{ 1\over 2}}
\def\thetaA {\theta_A}
\def\thetaB {\theta_B}

\def\phiA {\phi^A}

\def\adot {\dot a}

\def\gammu {{\gamma^\mu}}
\def\gamnu {{\gamma^\nu}}
\def\thetaB {\theta_B}
\def\PsiA {\Psi_A}

\def\grad {\partial}

\def\Gammu {\Gamma^\mu}

Discussions of the mechanics of particles with spin shows that
these can be described by either a particle theory with local world-line
supersymmetry [\Ref{one}], or
by a local fermionic symmetry [\Ref{two}].
This was generalised to superpace, to obtain a number of
{\it spinning} superparticle theories satisfying certain constraints
whose spectrum were precisely those of the ten-dimensional supersymmetric
Yang-Mills theory.
The quantum mechanics of a free superparticle in a ten-dimensional
space-time is of interest because of its close relationship to
ten-dimensional super Yang-Mills theory, and
this corresponds to the massless sector of type I superstring theory.
The $SO(9,1)$ covariant superfield formulation of super Yang-Mills theory
which reduces to $SO(8)$ formulation can be obtained by either an $SO(9,1)$
vector or spinor superfields. The superfields are chosen to satisfy either
rotational quadratic ( \lq\lq Type I")
or linear (\lq\lq Type II") constraints that restricts their field content
to the physical propagating fields.
The constraints are imposed by an explicit projection operator, constructed
out of super-covariant derivatives, acting on unconstrained superfields.
They were explicitly given on its $SO(8)$ form in [\Ref{brink}],
and presented on its $SO(9,1)$ form in [\Ref{two}].
Although, it is well known the difficult covariant
quantisation of superparticle
models due to the large number of symmetries that
they present[\Ref{prob}],
there are
by now several formulations which can be covariantly quantized.
These superparticle theories with spectra coinciding with
that of the ten dimensional super Yang-Mills are constructed by adding
appropriate
Lagrange multiplier terms to certain superparticle actions,
some of them leading to type I or II constraints.

This letter compares the BRST spectrum of type I and II superparticle
models arising from using the BV covariant quantisation
techniques[\Ref{bv}].
The zero ghost-number BRST cohomology class of these models gives the same
physical spectrum as that of D=10, N=1 super-Yang-Mills theory.
It is also shown that both quantum actions leads to free theories.
%
%
We begin by briefly reviewing the description of ten-dimensional
type I superparticle models [\Ref{twomike}].
A spinor wavefunction
can be obtained either from a spinning particle with local world-line
supersymmetry, or from a particle action with local fermionic symmetry.
In [\Ref{two}], it was seen that super Yang-Mills theory in ten-dimensions
is described by precisely such wavefunctions subject to certain extra
super-covariant constraints.
The quantum mechanics of the type I superparticle theory was
given in [\Ref{twomike}]. This superparticle action is formulated in an
extended ten-dimensional
superspace with coordinates $(\xmu ,\thetaA ,\phiA )$ where
$\thetaA$ and $\phiA$ are anti-commuting Majorana-Weyl spinors
\footnote
{$^\star$}
{A Majorana spinor $\Psi$ corresponds to a pair of Majorana-Weyl
spinors, $\PsiA$ and $\Psi^A$. The $32 \times 32$ matrices $C\gammu$
(where $C$ is the charge conjugation matrix) are block diagonal with
$16 \times 16$ blocks $\gammu^{AB}$, $\gammu_{AB}$ which are symmetric
and satisfy
$\gammu^{AB}\gamnu_{BC} + \gamnu^{AB}\gammu_{BC} =2\eta^{\mu\nu}\delta^A_C$.
In this notation the supercoordinates has components $\thetaA$,
$\theta\gammu\dot\theta = \thetaA \gammu_{AB} \dot\thetaB$,
$\psh_{AB} = p^\mu \gammu_{AB}$, etc. }.
The action is given by adding
$$
S_0 =\int\! d\tau \Big[ \pmu\dot\xmu +i\hat\theta\dot\theta
                         + i\hat\phi\dot\phi\Big],
\nfr{snutI}
and
$$
S_{1} =\int\! d\tau \Bigl[ -\half e p^2 + i\psi\psh d + i \Lambda\psh\hat\phi
          + \half( d \chi d - 2 \hat\phi\Gammu \chi \Gamma_\mu\psh\phi )
          +\half\hat\phi\Upsilon\hat\phi
\Bigr]\nfr{sbiprimI}
where
$\pmu$ is the momentum conjugate to the space-time coordinate $\xmu$, while
$\hat\theta =d -\psh\theta$ and $\hat\phi$ are the corresponding momentum
conjugate to the spinor coordinates $\theta$ and $\phi$, respectively.
The gauge fields $e$, $\psi$, $\Lambda$, $\chi$ and $\Upsilon$ correspond
to local symmetries, and impose the following constraints
$$
\eqalign{ p^2 =&0, \qquad \psh d =0, \qquad \psh\hat\phi =0, \cr
         \hat\phi\hat\phi &=0, \qquad
          d d + 2  \hat\phi\Gamma \Gamma \psh\phi =0.
\cr}\nfr{enconstry}
The physical
states are described by a superspace wavefunction
satisfying  $SO(9,1)$ covariant constraints[\Ref{two}]
$$\eqalign{
p^2 \Psi_A &=0, \qquad \psh_{AC}D^C \Psi_B =0, \qquad \psh^{AB}\Psi_B =0, \cr
&D^AD^B\Psi_C + 8 (\Gamma^\mu)^{E[A}(\Gamma_\mu\psh )^{B]}_{\ \ C}\Psi_E =0.
\cr}\nfr{quadra}
which leaves a superfield $\Psi_a (x^i,\theta^{\adot} )$
satisfying a quadratic projection condition
and describes precisely the $SO(8)$ constraints of ten dimensional
super Yang-Mills theory.
The covariant quantisation  of this superparticle was briefly
discussed in [\Ref{twomike}] in the gauge $e=1$ with the other gauge
fields set to zero. Covariant quantisation requires the methods of
Batalin and Vilkovisky [\Ref{bv}] since the gauge algebra only
closes on shell, and requires an infinite number of ghost fields
as the symmetries are infinitely reducible.
Following the BV formalism, it leads to a gauge-fixed quantum action
which, after field redefinitions and integrating out all non-propagating
fields, takes the form [\Ref{twomike}]
$$
S_Q =\int\! d\tau \Big[ \pmu\dot\xmu -\half p^2 + i\hat\theta\dot\theta
                       +i\hat\phi\dot\phi +\hat c\dot c +\hat\kappa\dot\kappa
                       +i\hat v\dot v + i\hat\rho\dot\rho
                       +\hat\zeta\dot\zeta \Big],
\nfr{quanty}
where
$\hat\theta = d - \psh\theta - 4\hat c\kappa$. This $S_Q$ proves to be
invariant under BRST transformations generated by the BRST charge
$$\eqalign{
Q =&\half c p^2 -Tr (\hat\rho\rho\psh\rho ) + \half i d\rho d
    + d\psh\kappa
    - 2\hat c\kappa\psh\kappa - 4 \hat c d\rho\kappa
    + 4i\hat c\hat\phi\Gammu\rho\Gamma_\mu \zeta\cr
    &+ 4i Tr (\hat v\Gammu\rho\Gamma_\mu\psh v)
    -2i \hat\phi\Gammu\rho\Gamma_\mu\psh\phi
    + \hat\phi\psh\zeta + \half i\hat\phi v\hat\phi .
\cr}
\nfr{chargeone}

The
quantum action defines a free field theory and so easy to quantize by
imposing canonical commutation relations on the operators corresponding
to the variables
$(\pmu ,\xmu ,e,\hat\theta ,\theta ,\hat\phi ,\phi ,\hat c ,c ,
\hat\kappa ,\kappa ,\hat v ,v ,\hat\rho ,\rho ,\hat\zeta ,\zeta )$.
It proves useful to choose a Fock space representation for the ghost
and define a ghost vaccum $|0>$ which is annihilated by each of the
antighosts
$(\hat\kappa |0> = 0,\hat c |0> = 0,\hat v |0> = 0,\hat\rho |0> =0,
  \hat\zeta |0> = 0 )$.
It also proves useful to define a {\it twisted} ghost vacuum $|0>_g$,
where for each ghost $g$ in the subscript, that ghost is an annihilation
operator and the corresponding anti-ghost is a creation operator.
The physical states on both twisted and untwisted Fock space
should be the same, as they are {\it dual} representations of
the same spectrum.
It is then viewed the superspace coordinates $\xmu$ ,$\hat\thetaA$ and
$\phiA$ as hermitian coordinates while $\pmu =-i\grad/\grad\xmu$,
$\hat\thetaA = \grad/\grad\thetaA$ and $\hat\phiA =\grad/\grad\phiA$.
A state of the form $\Phi (x,\theta ,\phi ) M|\Omega >$
with super-wavefunction $\Phi$, a monomial $M$ and $|\Omega >$ is considered.
The zero ghost-number cohomology class is given by $\Phi$, while $M$ is
constructed out of ghost and anti-ghost fields and $|\Omega >$
is one of the ghost ground state.
It was found then that the ghost-independent state
$\Phi (x,\theta ,\phi )|0>$ gives the physical spectrum
consisting of eight bosons and eight fermions which
form the $D=10$ super Yang-Mills multiplet together with the zero-momentum
ground state which is a supersymmetry singlet.

The general state of zero ghost-number and momentum $\pmu$ which is annhilated
by the BRST charge \chargeone\ satisfies the following conditions
$$\eqalign{
   p^2 \Phi &= 0, \qquad   \psh d\Phi = 0,
    \qquad \psh\hat\phi\Phi = 0, \cr
   \hat\phi\hat\phi \;\Phi &= 0, \qquad
   (d\; d - 8\hat\phi\Gammu\Gamma_\mu\psh\phi )\;\Phi =0. \cr}
\nfr{condiesone}
These
conditions \condiesone\
imply that the only non-vanishing parts of $\Phi (x,\theta ,\phi )$
are $\Psi_0 (x,\theta )$ and $\Psi_A (x,\theta )$. However, $\Psi_0$
is trivial unless $\pmu =0$. Thus the only non trivial part of $\Phi$
is $\Psi_A$ and satisfies precisely the covariant constraints \quadra ,
which lead to the D=10, N=1 super-Yang-Mills multiplet.
The monomial $M$ is constructed out of ghost and anti-ghost, and
to satisfy $Q^2 =0$, states of non-trivial ghost dependence should
include terms that involve $\Gammu\rho\Gamma_\mu$, which after
removing the bispinor parameters lead to the identity of $\Gamma 's$
to cancel in D=10, like in super-Yang-Mills theories appears
a term proportional to $\epsilon\psi^3$ which vanishes if supersymmetry
is to hold [\Ref{mbg}].

I shall now describe a ten-dimensional type II superparticle
model with a spinor super-wavefunction satisfying the following $S0(9,1)$
covariant linear constraints
$$\eqalign{
p^2 \Psi_A &=0, \qquad \psh_{AC}D^C \Psi_B =0, \qquad \psh^{AB}\Psi_B =0, \cr
&(\Gamma^{\mu\nu\rho\sigma})_A^{\ \ B} D^A\Psi_B =0,
\cr}\nfr{linear}
which
proves to be equivalent to constraints \quadra .
This model is also formulated in an extended ten-dimensional
superspace with coordinates
$(\xmu ,\thetaA ,\phiA )$ where $\thetaA$ and $\phiA$ are
anticommuting Majorana-Weyl spinors, and to describe super Yang-Mills
we wish to impose the extra constraints
$d^A (\Gamma^{\mu\nu\rho\sigma})^B_{\ A} =0$ and $\hat\phi\hat\phi =0$
which can be done
by adding appropriate lagrange multiplier terms ( See Ref. \Ref{two}
for further details ).

The type II superparticle action is then given by the
sum of [\Ref{two}]
$$
S_0 =\int\! d\tau \Big[ \pmu\dot\xmu +i\hat\theta\dot\theta
                         + i\hat\phi\dot\phi\Big],
\nfr{snut}
and
$$
S_{1} =\int\! d\tau \Big[ -\half e p^2 + i\psi\psh d
           + i \varphi\psh\hat\phi  + i d\Lamdsh\hat\phi
           - i\beta(\phi\hat\phi - 1) + \half\hat\phi\omega\hat\phi
           \Big] \nfr{sbiprim}
where,
as usual,
$\pmu$ is the momentum conjugate to the space-time coordinate $\xmu$,
$d^A$ is a spinor introduced so that the Grassmann coordinate $\thetaA$
has a conjugate momentum $\hat\theta^A =d^A -\psh^{AB}\thetaB$,
$\phiA$ is a new spinor coordinate and $\hat\phi_A$ is its conjugate
momentum.
The fields $e$, $\psi^A$, $\varphi_A$, $\Lambda_{\mu\nu\rho\sigma}$,
$\beta$ and $\omega^{AB}=-\omega^{BA}$ are all
Lagrange multipliers, which are
also gauge fields for corresponding local symmetries and impose the
following constraints
$$
\eqalign{ p^2 =&0, \qquad \psh d =0, \qquad \psh\hat\phi =0,\cr
         \hat\phi\hat\phi =0, \qquad
         &\phi\hat\phi - 1 =0, \qquad
         d(\Gamma^{\mu\nu\rho\sigma})\hat\phi =0.
\cr}\nfr{constry}
Physical
states are described by a superspace wavefunction
satisfying  $SO(9,1)$ covariant linear constraints \linear ,
which leaves a superfield $\Psi_a (x^i,\theta^{\adot} )$
satisfying a linear projection condition
which is precisely the $SO(8)$ constraint of the ten dimensional
super Yang-Mills theory.
The free quantum action is given by
$$
S_Q =\int\! d\tau \Big\{\pmu\dot\xmu + i\hat\theta\dot\theta
       +i\hat\phi\dot\phi +\half p^2 +\hat c\dot c
       +\hat\kappa\dot\kappa +\hat\zeta\dot\zeta
       +\hat\Sigmash\dot\Sigmash +\hat\eta\dot\eta
       +\hat v\dot v \Big\}.
\nfr{quadryx}
where
$\hat\theta = d-\psh\theta -4i\hat c\kappa_1$. This quantum action is invariant
under modified BRST transformations
which are generated by
the following conserved $(\dot Q_{BRST}=0)$ and nilpotent
$(Q^2_{BRST} =0)$ BRST charge
$$\eqalign {
 Q_{BRST} = &\half c p^2 + 2 i d\psh\kappa + 2 i \hat\phi\psh\zeta
            - 2 i d\hat\phi\Sigmash -\hat\phi v\hat\phi -\hat\phi\eta\phi
            + \hat\zeta\zeta\eta +\hat\Sigmash\Sigmash\eta \cr
          & +\hat\eta\eta\eta
            + 2 \hat v v\eta - 2 \hat c\hat\theta\kappa_2
            - 2 i \hat c\kappa\psh\kappa + i \hat\kappa\psh\kappa_2
            + 4 \hat c\hat\zeta\kappa_2\Sigmash
            + 2 i \hat\kappa\hat c\kappa_3
            + 2 i \hat v\Sigmash\psh\Sigmash .
\cr}\nfr{brstcharge}
The
quantum
action \quadryx\ is free and the superparticle type II can again be
quantized canonically by replacing each of the fields by an operator
and imposing canonical (anti-) commutation relations on the conjugate pairs
$(\pmu ,\xmu )$, $(\hat\theta ,\theta )$, $(\hat\phi ,\phi)$,
$(\hat c, c)$, $(\hat\kappa_1 ,\kappa_1)$, $(\hat\zeta_1 ,\zeta_1)$,
$(\hat\Sigmash_1 ,\Sigmash_1)$, $(\hat\eta_1 ,\eta_1)$ and
$(\hat v_1, v_1)$.
The
cohomology classes can be classified according
to their total ghost number and the physical states are taken to be
the cohomology class of some definite ghost number.
We consider
two distinct Fock space representations of the ghost system,
the {\it untwisted} one in which the ghost ground state $|0>$ is
annihilated by each of the antighosts
$(\hat\kappa_1 |0> = 0,\hat c |0> = 0,\hat v_1 |0> = 0,\hat\eta_1 |0> =0,
  \hat\Sigmash_1 |0>=0, \hat\zeta_1 |0> = 0 )$,
and the {\it twisted}
one in which antighosts are creation operators and ghosts are
annihilation operators.
It is viewed $x$, $\theta$ and $\phi$ as hermitian coordinates while
$\pmu =-i\grad/\grad\xmu$, $\hat\thetaA = \grad/\grad\thetaA$ and
$\hat\phiA =\grad/\grad\phiA$, and
consider again states of the form $\Phi (x,\theta ,\phi )\;M\;|\Omega >$
with super-wave function $\Phi$. $M$ is some monomial constructed out of
ghosts and anti-ghost and $|\Omega >$ is one of the ghost ground state.
A general state
of zero-ghost number and momentum $\pmu$ which is annhilated by the
BRST charge \brstcharge\ satisfies
$$\eqalign { &
    p^2 \Phi =0, \qquad \psh d\Phi =0, \qquad \psh\hat\phi\Phi =0,  \cr
   \hat\phi\hat\phi\Phi &=0, \qquad (\phi\hat\phi -1)\Phi=0, \qquad
   d\Gamma^{\mu\nu\rho\sigma}\hat\phi\Phi =0.
\cr}\nfr{constries}
These
implies that the only non-vanishing parts of $\Phi (x,\theta,\phi )$
are $\Psi_0 (x,\theta )$ and $\Psi_A (x,\theta )$. However,
$(\phi\hat\phi -1)\Phi = 0$ implies that $\Psi_0$ is trivial.
The only non-trivial part of the super-wave function is
$\Psi_A$ which precisely
satisfies the covariant constraints \constry\ leading
to the D=10, $N=1$ super-Yang-Mills multiplet.
The monomial
$M$ is constructed out of ghost and anti-ghosts and to satisfy
$Q^2=0$, states of non-trivial ghost dependence include terms that
involve products of $\Gamma 's$, like the term $\Sigmash\psh\Sigmash$
which after removing the multi-tensor bosonic parameters leads
also to an identity of $\Gamma 's$ in $D=10$, and needed to avoid the
propagation of these bosonic ghost fields.
Lets recall that the covariant quantisation of the $SSP2$ superparticle model
lead to the appearance of multispinor ghosts and rouse the status
of the BRST cohomology of that model [\Ref{newyork}].

\sjump

{\it acknowledgments}: I wish to thank J.A. Helayel-Neto for reading and
comments of the manuscript.

\bjump

\eject

\centerline{\capsone REFERENCES}

\sjump
\beginref
\Rref{one}{L. Brink, S. Deser, B. Zumino, P. DiVecchia and P.S. Howe,
Phys. Lett. {\bf B64} (1976) 435; F.A. Berezin and M.S. Marinov,
Ann. Phys. {\bf 104} (1977) 336; L. Brink, P. DiVecchia and P.S. Howe,
Nucl. Phys. {\bf B118} (1977) 76; C. Galvao and T. Teitelboim,
J. Math. Phys. {\bf 21} (1980) 1863; P.S. Howe, S. Penati; M. Pernici and
P.K. Townsend, Phys. Lett. {\bf 215} (1988) 555;
V.D. Gershum and V.I. Tkach, JETP Lett. {\bf 29} (1979) 320;
W. Siegel, Int. J. Mod. Phys. {\bf A3} (1988) 2713.}
\Rref{two}{C.M.Hull and J.L. V\'azquez-Bello,
Nucl. Phys. B416 (1994) 173.}
\Rref{twomike}{M.B. Green and C.M. Hull, Mod. Phys. Lett. A18 (1990)
1399.}
\Rref{brink}{L. Brink, M.B. Green and J.H. Schwarz, Nucl. Phys.
{\bf B223} (1983) 125.}
\Rref{prob}{W. Siegel, Phys. Lett. {\bf B203} (1988) 79;
L. Brink, in {\it Physics and Mathematics of Strings}, eds. L. Brink, D.
Friedan and A.M. Polyakov, (World Scientific, 1990);
A. Mikovi\'c, M.Ro\v cek, W. Siegel, A.E. van de Ven, P. van Nieuwenhuizen
and J.P. Yamron, Phys. Lett. {\bf B235} (1990) 106;
F. E$\beta$ler, E. Laenen, W. Siegel and J.P. Yamron, Phys. Lett. {\bf B254}
(1991) 411; F. E$\beta$ler, M. Hatsuda, E. Laenen, W. Siegel, J.P. Yamron,
T. Kimura and A. Mokovi\'c, Nucl. Phys. {\bf B364} (1991) 67;
E.A. Bergshoeff and R.E. Kallosh, Phys. Lett. {\bf B240} (1990) 105;
Nucl. Phys. {\bf B333} (1990) 605; E.A. Berghsoeff, R.E. Kallosh and
A. van Proeyen, Phys. Lett. {\bf B251} (1990) 128;R.E. Kallosh,
Phys. Lett. {\bf B251} (1990) 134; M. Huq, Int. J. Mod. Phys. {\bf A7}
(1992) 4053; J.L. V\'azquez-Bello, Int. J. of Mod. Phys.
{\bf A19} (1992) 4583.}
\Rref{bv}{I.A. Batalin and G.A. Vilkovisky, Phys. Lett. {\bf B102}
(1981) 462;Phys. Rev. {\bf D28} (1983) 2567.}
\Rref{twomike}{M.B. Green and C.M. Hull, Mod. Phys. Lett. A18 (1990)
1399.}
\Rref{mbg}{ M.B. Green, J. H. Schwarz and E. Witten, {\it
Superstrings} Eds. Cambridge Univerty Press (1987).}
\Rref{newyork}{U. Lindstr\" om, M Ro\v cek, W. Siegel, P. Van Nieuwenhuizen and
A. E. Van de Ven, J. Math. Phys. {\bf 31} (1990) 1761.}

\endref
\ciao